\documentstyle[preprint,aps,eqsecnum]{revtex}
\begin{document}
\draft
\preprint{IP/BBSR/94-19}
\title
{\bf Vacuum Structure and Chiral Symmetry Breaking }
\author{S.P. Misra}
\address
{Institute of Physics, Bhubaneswar-751005, India.}
\maketitle
\begin{abstract}
We relate here the vacuum structure of chiral symmetry breaking to a wide
class of low energy hadronic properties to examine consistency with
experimental results. Phase transition of chiral symmetry breaking is
described through explicit vacuum realignment. We find that for the present
ansatz, vacuum structure of the light quark sector can be determined
from $f_\pi=92$ MeV, which then yields that $\Gamma(\pi^0\rightarrow
2 \gamma)=8.3$ eV and, $R_{ch}=0.67$ fms, in agreement with
experiments. This amounts to three independent `observations' of the vacuum
structure of the light quark sector. For the calculation of
$\Gamma(\pi^0\rightarrow 2\gamma)$,
we do not use the anomaly equation - local gauge invariance for
vacuum realignment becomes adequate. The above width is sensitive to
the vacuum structure, and can give evidence regarding progress towards
chiral symmetry restoration in QGP. The corresponding results
for $s$ quark sector are moderately good. We also exploit similar
ideas for heavy quarks, but here the experimental data are rare, and the
theoretical techniques are likely to need refinements.
\end{abstract}
\newpage
\def\zbf#1{{\bf {#1}}}
\def\bfm#1{\mbox{\boldmath $#1$}}
\def\hf{\frac{1}{2}}

\section {\bf Introduction}
It is now accepted that quantum chromodynamics (QCD) is the correct theory
for strong interaction physics of quarks and gluons, and, at a secondary
level, of hadrons. However, at present no reliable method
is known for the solution of such problems. The basic difficulty appears to
be an understanding of the ground state properties of QCD or its vacuum
structure, which plays an important role \cite{ltg}, and has been
under intense investigation since a long time dealing with the nontrivial
vacuum structure for gluons \cite{qcd,svz,amqcd}.
A parallel nonperturbative approach for quark condensates has also
been made through chiral symmetry breaking with pion as a Goldstone mode
\cite{chrl}. We have looked at the same in
the context of potentials \cite{amcrl} as well as for Nambu Jona Lasinio model
\cite{hmnj} with symmetry breaking as vacuum realignment \cite{spmtlk}.

Our approach to these problems \cite{amqcd,amcrl,hmnj} mainly has been
variational, where the parameters or functions are determined through
minimisation of energy density. At the present level of sophistication,
determination of the vacuum structure through such a
minimisation does not seem to be very reliable. In view of this we
had also considered an alternative approach \cite{amspm} where on intuitive
grounds we take a simple ansatz for the vacuum structure, and then
examine the consequences for observed hadronic properties. This appeared
to describe a host of phenomena as being related to the vacuum structure
for chiral symmetry breaking \cite{amspm}. Here we shall generalise the same
to discuss a wider class of problems to determine the vacuum structure
from experimental observations.

The paper is organised as follows. In section {\bf II} we discuss an ansatz
for the vacuum structure for chiral symmetry breaking to see whether {\it
post facto} it can be the correct description for vacuum realignment.
In particular we obtain the four component Dirac spinors as related to the
vacuum structure. In section {\bf III} we obtain the pion and kaon wave
functions in terms of vacuum realignment of $u$, $d$ and $s$ quarks. We then
use the experimental values of $f_\pi$ and $f_K$ {\it to determine}
the vacuum realignment of the above quarks for the present ansatz.
In section {\bf IV} we consider the process $\pi^0\rightarrow 2\gamma$
as a consequence of vacuum realignment with local gauge invariance.
This yields that $\Gamma(\pi^0\rightarrow 2\gamma)=8.3$ eV {\it without}
fixing any additional parameter or using the anomaly equation.
In section {\bf V} we discuss the masses of the pseudoscalar mesons through
approximate chiral symmetry breaking in terms of the `small'
Lagrangian masses of the quarks. We then use the same to
determine the vacuum structure for $c$-quark and $b$-quark condensates,
and obtain their decay constants along with some additional results.
In section {\bf VI} we obtain the charge radii
of the pion and the kaon where the mesons in motion are defined through
Lorentz boosting of the corresponding states at rest.
In section {\bf VII} we summarise the results, mention the limitations
of the present paper and discuss the scope for further work. We also
discuss here the relevance of the present ideas in quark gluon plasma
for chiral symmetry restoration.

The method considered here
is a non-perturbative one as we use only equal time quantum algebra
but is limited by the choice of ansatz functions in the calculations.
The techniques have been applied earlier to solvable cases to examine
its reliability \cite{spmbg} as well as to physically more relevant ground
state structures in high energy physics \cite{amqcd,higgs} and nuclear physics
\cite{nmtr,dtrn}. We examine here chiral symmetry breaking with an ansatz,
and then relate the same to low energy hadronic properties in an attempt
to identify vacuum structure through experimental results.

\section{ chiral symmetry breaking for quarks}
For the consideration of chiral symmetry breaking, we shall take the
perturbative vacuum state with chiral symmetry as $|vac>$. In this basis
quarks are massless, or have a small mass. We shall then
assume a specific vacuum realignment which breaks chiral symmetry.
As stated we shall then relate the vacuum structure with experimental
observations \cite{amspm}.

We have seen earlier \cite{amcrl,hmnj,spmtlk,amspm} that chiral symmetry
breaking takes place with the formation of quark antiquark condensates in
the perturbative vacuum. We shall thus take the destabilised vacuum,
$|vac'>$ as
\begin{equation}
|vac'>=U_Q|vac>
\label{u0}
\end{equation}
where $U_Q$ is given as
\begin{equation}
U_Q=e^{B_0 ^\dagger-B_0}
\label{u1}
\end{equation}
with
\begin{equation}
B_0 ^\dagger=\int q^{0i}_I(\zbf k)^\dagger h_i(\zbf k)
(\bfm\sigma\cdot\hat\zbf k)\tilde q^{0i}_I(-\zbf k)d\zbf k.
\label{b0}
\end{equation}
In the above, $i$ stands for the flavour of the quark with $i=1,2,3,\cdots$
standing for $u,d,s,\cdots$ quarks. The subscript $I$ and the superscript
`0' indicate that these are two component quark or antiquark operators
which create or annihilate quanta of the perturbative or chiral vacuum.
We are accepting the fact that there may be condensates also for heavy
quarks. $h^i(\zbf k)$ are the ansatz functions which describe vacuum
realignment for quark of flavour $i$. We shall take a simple form for
them on intuitive grounds \cite{amspm}
and then relate the same to observations. The colour quantum number of
quarks has been suppressed. We are taking the quarks after a Kobayashi
Maskawa rotation \cite{donogh}, so that the Lagrangian mass matrix for them is
diagonal. Our ansatz here has an obvious parallel with
superconductivity \cite{chrl,anderson}.

The above two component creation or annihilation operators arise
in the momentum space expansion of the four component field operators
corresponding to the perturbative basis. They are given as \cite
{amcrl,hmnj,spmtlk,amspm}
\begin{eqnarray}
\psi (\zbf x )\equiv &&\frac{1}{(2\pi)^{3/2}}\int \tilde\psi(\zbf k)
e^{i\zbf k\cdot\zbf x}d\zbf k \nonumber\\
=&&\frac{1}{(2\pi)^{3/2}}\int \left[U_0(\zbf k)q^0_I(\zbf k )
+V_0(-\zbf k)\tilde q^0_I(-\zbf k )\right]e^{i\zbf k\cdot \zbf x}d \zbf k,
\label{psiexp}
\end{eqnarray}
where \cite{spm78,lopr}
\begin{eqnarray}
U_0(\zbf k )=&&\left(\begin{array}{c}cos\frac{\chi_0(\zbf k)}{2}\\
\zbf \sigma \cdot \hat k sin\frac{\chi_0(\zbf k)}{2}\end{array}\right),
\nonumber\\
V_0(-\zbf k )=&&\left( \begin{array}{c} -\zbf \sigma \cdot \hat k
sin\frac{\chi_0(\zbf k)}{2} \\ cos\frac{\chi_0(\zbf k)}{2}\end{array}
\right).
\label{uv0}
\end{eqnarray}
The above form is so taken that it satisfies the equal time algebra
\cite{spm78,lopr}, and we have suppressed both colour and flavour indices.
In the above, for {\it free} quark fields of mass $m_Q$, we have \cite{spm78}
\begin{equation}
cos\frac{\chi_0(\zbf k)}{2}=\Big({p^0+m_Q \over 2p^0}\Big)^{1 \over 2},
\qquad
sin\frac{\chi_0(\zbf k)}{2}=\Big({p^0-m_Q \over 2p^0}\Big)^{1/2},
\label{chi0}
\end{equation}
with $p^0=\big(m^2_Q+\zbf k^2 \big)^{1 \over 2}$.
We may further note that {\it when chiral symmetry is unbroken and thus
the quarks are massless}, $\chi_0(\zbf k)=\pm\pi/2$. As we shall now see,
this can be used in an essential manner to obtain the four component
quark field operators when chiral symmetry gets broken.

Now, for any operator $O$ in the $|vac>$ basis, the corresponding operator
$O'$ in the $|vac'>$ basis is given through
\begin{equation}
O'=U_QO U_Q ^\dagger
\label{oop}
\end{equation}
which allows us to write the creation and annihilation operators in the
$|vac'\rangle $ basis. We may here first note from equation (\ref{oop}) that
{\it the operator $U_Q$ remains unchanged in form when we go from one basis to
the other}, which follows from the fact that $U_Q'=U_QU_QU_Q ^\dagger=U_Q$.
We can thus take the operator $U_Q$ as defined by equations (\ref{u1})
and (\ref{b0}) in $|vac\rangle$ basis, or in $|vac'\rangle$ basis.

Now, the creation and annihilation operators in the chiral symmetry
broken basis are given through the inverse of equation (\ref{oop}) as
\begin{eqnarray}
&&\left(\begin{array}{c}q^0_I(\zbf k)\\
\tilde q^0_I(-\zbf k)\end{array} \right)=
U_Q ^\dagger \left(\begin{array}{c}q_I(\zbf k)\\
\tilde q_I(-\zbf k)\end{array} \right)U_Q \nonumber\\
=&&
\left(\begin{array}{cc}cos(h(\zbf k))& \zbf \sigma\cdot\hat k sin(h(\zbf k))\\
-\zbf \sigma\cdot\hat k sin(h(\zbf k))& cos(h(\zbf k))\end{array} \right)
\left(\begin{array}{c}q_I(\zbf k)\\
\tilde q_I(-\zbf k)\end{array} \right).
\label{bgt}
\end{eqnarray}
In the above we have taken $U_Q$ in the $|vac'\rangle$ basis, with
$q_I(\zbf k)$ and $\tilde q_I(-\zbf k)$ also being the two-component
quark annihilation and antiquark creation operators in $|vac'\rangle$.
As earlier we have suppressed the spin, flavour and colour
indices. From equations (\ref{psiexp}) and (\ref{bgt}) we relate the
quark field expansions in the two basis through the equation
\begin{equation}
\tilde\psi(\zbf k)=U_0(\zbf k)q^0_I(\zbf k )
+V_0(-\zbf k)\tilde q^0_I(-\zbf k )
=U(\zbf k)q_I(\zbf k )+V(-\zbf k)\tilde q_I(-\zbf k )
\label{psiop}
\end{equation}
so that the four component spinors after chiral symmetry breaking become
known. We then easily obtain from equation (\ref{bgt}) that
the spinors $U(\zbf k)$ and $V(\zbf k)$ after chiral symmetry breaking are
given as
\begin{eqnarray}
U(\zbf k )=&&\left(\begin{array}{c} cos\frac{\chi(\zbf k)}{2} \\
\zbf \sigma \cdot \hat k sin\frac{\chi(\zbf k)}{2}\end{array}\right),
\nonumber\\ \nonumber\\
V(-\zbf k )=&&\left(\begin{array}{c} -\zbf \sigma \cdot \hat k
sin\frac{\chi(\zbf k)}{2} \\ cos\frac{\chi(\zbf k)}{2}\end{array}\right),
\label{uv}
\end{eqnarray}
where,
\begin{equation}
\frac{\chi(\zbf k)}{2}=\frac{\chi_0}{2}-h(\zbf k).
\end{equation}
{\it Now, when we know $\chi_0$ we also know $\chi(\zbf k)$ in terms of
$h(\zbf k)$ for the vacuum structure}.
For definiteness we take for the chiral vacuum $\chi_0=\pi/2$.
{\it Then the vacuum structure explicitly gives also the four component quark
field operators} \cite{amspm}. In what follows we shall exploit this result.
We also note that $f_i(\zbf k)$ and $g_i(\zbf k)$
in Ref. \cite{spm78} for quarks of flavour $i$ here correspond to
\begin{equation}
f_i(\zbf k)=cos\left(\hf\chi_i(\zbf k)\right);\qquad
|\zbf k|g_i(\zbf k)=sin\left(\hf\chi_i(\zbf k)\right).
\label{fqgq}
\end{equation}

We further note that
\begin{equation}
<vac'|\bar\psi_i(\zbf x)\psi_i(\zbf y)|vac'>
=-\frac{6}{(2\pi)^3}\int cos\chi_i(\zbf k)e^{i\zbf k\cdot(\zbf x -
\zbf y)}d \zbf k,
\label{expqq}
\end{equation}
where, as earlier $i$ is the flavour of the quark, and, the factor 6 arises
from colour and spin degress of freedom. We note that in the above
$cos\chi_i(\zbf k)=
sin(2h_i(\zbf k))$, describing the above function in terms of
the quark antiquark correlations for vacuum realignment.

We now make our intuitive approximation about the vacuum structure on the
basis of the above results. {\it We shall approximate the function
in equation (\ref{expqq}) by a Gaussian}. Thus, with $cos\chi_i(\zbf 0)=1$
in equation (\ref{uv}), we write
\begin{equation}
sin(2h_i(\zbf k))=cos(\chi_i(\zbf k))=e^{-\hf R_i^2\zbf k^2},
\label{sin2h}
\end{equation}
with a single dimensional parameter $R_i$ for each quark of flavour $i$.
This simple form appears to be consistent with some
low energy hadronic phenomenology as earlier \cite{amspm} and
we shall take the same as a ``zeroeth order" ansatz.

We shall further assume that initially we have $SU(2)_L\times SU(2)_R$ chiral
symmetry which breaks to the custodial symmetry $SU(2)_V$. We then
have $R=R_1=R_2$ for $u,d$ quarks.

\section{Vacuum structure and pion and kaon decay constants}
In the present section we shall first use Goldstone theorem to identify
the quark antiquark wave functions for the $\pi^+$ and $K^+$ states
as related to the vacuum structure \cite{amcrl,hmnj,amspm}, and then
link the same to the corresponding decay constants. For the sake of
completeness, we briefly recall \cite{amspm} the calculations for pion
here.

We note that the chiral generator corresponding to the $\pi^+$ state is
given as
\begin{equation}
Q_5^{\pi^+}=\int\psi_1(\zbf x)^\dagger \gamma_5\psi_2(\zbf x)d\zbf x,
\label{pipgen}
\end{equation}
with $\psi_1$ and $\psi_2$ standing for $u$ and $d$ quarks.
Clearly when chiral symmetry is there, $Q_5^{\pi^+}|vac\rangle=0$,
whereas when chiral symmetry is broken, $Q_5^{\pi^+}|vac'\rangle\ne 0$,
and will correspond to the $\pi^+$ state of zero momentum
\cite{amcrl,hmnj,spmtlk,amspm}.
Hence on using equation (\ref{pipgen}), through a direct evaluation we
identify the $\pi^+$ state of zero momentum as, with
$cos\chi(\zbf k)=sin(2h(\zbf k))$,
\begin{equation}
\mid \pi^+(\zbf 0)>=N_{\pi}\cdot\frac{1}{\sqrt6}\int u_I(\zbf k)^\dagger
{\tilde d_I}(-\zbf k)cos(\chi(\zbf k))
d\zbf k \mid vac'>,
\label{pipst}
\end{equation}
where, $N_\pi$ is a normalisation constant. For the $u$ and $d$
quarks we have $R_1=R_2=R$, and thus $\chi_1(\zbf k)=
\chi_2(\zbf k)\equiv \chi(\zbf k)$. The colour and spin indices of
quarks and antiquarks have been suppressed. The normalisation
constant $N_\pi$  is given through the equation \cite{spm78}
\begin{equation}
{N_{\pi}}^2 \int cos^2(\chi(\zbf k)) d\zbf k =1.
\label{norm1}
\end{equation}
For the present ansatz of equation (\ref{sin2h}) we then have from
equation (\ref{pipst}) the pion wave function $\tilde u_\pi(\zbf k)$ as
\cite{spm78}
\begin{equation}
\tilde u_{\pi}(\zbf k)=N_{\pi}cos(\chi(\zbf k))=N_{\pi}\exp^{-\hf
R^2\zbf k^2},
\label{piwvf}
\end{equation}
where, on integration we have
\begin{equation}
N_{\pi}=\frac{R_{\pi}^{3/2}}{\pi^{3/4}} =0.424\times R_{\pi}^{3/2}.
\label{pinorm}
\end{equation}

Clearly the state as in equation (\ref{pipst}) as the Goldstone mode will be
accurate to the extent we determine the vacuum structure sufficiently
accurately through variational or any other method so that $|vac'>$ is
an eigenstate of the total Hamiltonian. In general, if $Q_a$ is a
generator for a symmetry breaking direction of a global symmetry
\cite{glstn}, then
\begin{equation}
Q_a|vac\rangle=0
\end{equation}
whereas
\begin{equation}
Q_a|vac'\rangle\ne 0,
\label{qivcp}
\end{equation}
and, it defines the state for the Goldstone mode
\cite{amcrl,hmnj,spmtlk,amspm}. This result
in particular yields wave function of the pion from the
vacuum structure for any example of chiral symmetry breaking,
and is the new feature of looking at phase transition
through vacuum realignment \cite{spmtlk}.

We had earlier considered low energy hadronic properties \cite{spm78}
with an assumed form of four component quark field operators. This is
equivalent to a choice of $cos(\chi(\zbf k))$. In addition, we had
taken Gaussian wave functions for mesons and then discussed
the phenomenology. Now from the vacuum structure the pion wave function
is known and, the four component quark field operators are also known.
This decreases the number of independent quantities making
the earlier model \cite{spm78} more restrictive, giving rise
to predictions.

We next use the pion decay constant $f_\pi$ to determine the vacuum
structure of chiral symmetry breaking for $u$ and $d$ quarks \cite{amspm}.
The decay constant $f_{\pi}$ as calculated earlier in terms of the
wave function including here the relativistic corrections
\cite{spm78,van} is given through
\begin{equation}
\left|(1+2g_1^2 \zbf \bigtriangledown ^2)u_{\pi}(\zbf 0)\right|
=\frac{f_{\pi}(m_{\pi})^{1/2}}{\sqrt 6},
\label{vanfpi}
\end{equation}
where the value of the wave function is taken at the space origin, and,
$g_1$ is a differentiation operator in coordinate space given through
equation (\ref{fqgq}) with $g_1=g_2$ for $u$ and $d$ quarks. On
taking the Fourier transform, the left hand side of the above equation
becomes
\[
\frac{1}{(2\pi)^{3/2}}\int cos\chi(\zbf k)u_{\pi}(\zbf k)d \zbf k,
\]
where we have substituted that $1-2g_1(\zbf k)^2\zbf k^2
=cos(\chi(\zbf k))$. Hence from equations (\ref{norm1}), (\ref{piwvf}) and
(\ref{vanfpi}) we obtain that
\begin{equation}
\frac{1}{(2\pi)^{3/2}}\cdot\frac{1}{N_{\pi}}
=\frac{f_{\pi}(m_{\pi})^{1/2}}{\sqrt 6},
\label{fpi}
\end{equation}
so that, with $f_\pi=92$ MeV, we get
\begin{equation}
N_{\pi}=4.534\;{\rm GeV}^{-3/2}.
\label{pinorm1}
\end{equation}
Hence, from equation (\ref{pinorm}) we obtain that, with $R_1=R_2$,
\begin{equation}
R_1^2=R_2^2=\pi\times N_{\pi}^{4/3}=23.58\;{\rm GeV}^{-2}.
\label{rpi2}
\end{equation}
This (i) determines the vacuum structure in the light quark sector as in
equation (\ref{sin2h}), (ii) yields the pion wave function as in equation
(\ref{piwvf}), and, (iii) gives the four component quark field operators for
light quarks through eqaution (\ref{uv}).

We next utilise the value of $f_K$ to determine the vacuum structure
for the $s$-quark. The generator $Q_5^{K^+}$ for $K^+$ is given as, with
$\psi_3$ standing for the $s$-quark,
\begin{equation}
Q_5^{K^+}=\int\psi_1(\zbf x)^\dagger \gamma_5\psi_3(\zbf x)d\zbf x.
\label{kpgen}
\end{equation}
Hence on direct evaluation we identify the $K^+$ state of zero momentum as
\begin{eqnarray}
\mid K^+(\zbf 0)>&&=N_{K}\cdot\frac{1}{\sqrt6}\int u_I(\zbf k)^\dagger
{\tilde s_I}(-\zbf k)\Big[cos(\hf\chi_1(\zbf k))cos(\hf\chi_3(\zbf k))
\nonumber\\ &&
-sin(\hf\chi_1(\zbf k))sin(\hf\chi_3(\zbf k))\Big]
d\zbf k \mid vac'>,
\label{kpst}
\end{eqnarray}
where, $N_K$ is a normalisation constant.
The wave function for kaon $\tilde u_{K}(\zbf k)$ thus is given as
\begin{equation}
\tilde u_K(\zbf k)=N_K\Big[cos(\hf\chi_1(\zbf k))cos(\hf\chi_3(\zbf k))
-sin(\hf\chi_1(\zbf k))sin(\hf\chi_3(\zbf k))\Big],
\label{kwvf}
\end{equation}
with the normalisation constant $N_K$  determined through
\begin{equation}
N_K^2 \int\Big[cos(\hf\chi_1(\zbf k))cos(\hf\chi_3(\zbf k))
-sin(\hf\chi_1(\zbf k))sin(\hf\chi_3(\zbf k))\Big]^2 d\zbf k =1,
\label{normk}
\end{equation}
which gives
\begin{equation}
\frac{1}{N_K^2}=2\pi\int\Big[1-sin(\chi_1(|\zbf k|))sin(\chi_3(|\zbf k|))
+cos(\chi_1(|\zbf k|))cos(\chi_3(|\zbf k|))\Big]|\zbf k|^2d|\zbf k|.
\end{equation}
Parallel to equation (\ref{vanfpi}), the Van Royen Weisskopf
relation for $K^+$ meson in momentum space is {\it now calculated as}
\cite{spm78}
\begin{equation}
\frac{1}{(2\pi)^{3/2}}
\left|\int\big[cos(\hf\chi_1(\zbf k))cos(\hf\chi_3(\zbf k))
-sin(\hf\chi_1(\zbf k))sin(\hf\chi_3(\zbf k))\big]
u_K(\zbf k)d\zbf k\right|
=\frac{f_K(m_K)^{1/2}}{\sqrt 6},
\label{vanfk}
\end{equation}
which from equations (\ref{kwvf}) and (\ref{normk}) is equivalent to
\begin{equation}
\frac{1}{(2\pi)^{3/2}}\cdot\frac{1}{N_K}
=\frac{f_K(m_K)^{1/2}}{\sqrt 6}.
\label{fk}
\end{equation}
With $f_K=1.2\;f_\pi$, this yields that
\begin{equation}
N_K=1.9935\;{\rm GeV}^{-3/2}.
\label{normk1}
\end{equation}
Now, in equation (\ref{normk}) $R_1$ is known as in equation (\ref{rpi2}).
Hence with equation (\ref{normk1}) we can obtain the value of $R_3=R_s$.
Through a numerical evaluation we then obtain that
\begin{equation}
R_s^2=4.084\;{\rm GeV}^{-2}.
\label{rs2}
\end{equation}
This determines the vacuum structure for the $s$-quark, the four component
spinors for the $s$-quark, as well as the kaon wave function
in the limit of exact chiral symmetry breaking.

We can now obtain also the values of the $u,\;d$ and $s$ quark antiquark
condensates. In fact,
from equations (\ref{expqq}) and (\ref{sin2h}) we easily obtain that
\begin{equation}
-\langle vac'|\bar \psi_i(\zbf x)\psi_i(\zbf x)|vac'\rangle
=\frac{3}{\sqrt 2}\cdot\frac{1}{\pi\sqrt\pi}\cdot\frac{1}{R_i^3},
\label{qcnd}
\end{equation}
so that we have
\begin{equation}
\big(-\langle\bar uu\rangle\big)^{1/3}
=149 {\rm\; MeV};\quad
\big(-\langle\bar u u\rangle-\langle\bar dd\rangle\big)^{1/3}
=188 {\rm\; MeV};\quad
\big(-\langle\bar ss\rangle\big)^{1/3}
=359 {\rm\; MeV}.
\label{udscnd}
\end{equation}
The condensate expressions for $u$ and $d$ quarks are
smaller than usual. Chiral symmetry is restored when $R_i\rightarrow
\infty$, and, the condensates as above vanish.

\section{minimal electromagnetic coupling and $\pi^0\rightarrow 2\gamma$ decay}
In this section we shall see that the vacuum structure in the light quark
sector along with local gauge invariance of $|vac'\rangle$ will be adequate
to yield the $\pi^0$ width. For this purpose we shall obtain minimal coupling
as explained below, and then calculate $\Gamma(\pi^0\rightarrow 2\gamma)$.
We shall see that this yields the correct experimental value of the same
{\it without} additional parameters, and without using the anomaly equation.

In order to include electromagnetic interactions correctly, we should
consider vacuum realignment as in equations (\ref{u1}) and (\ref{b0}) so
that gauge invariance is maintained while restructuring of vacuum takes place.
Let us consider the covariant derivative of the quark field $\psi_q(\zbf x)$
given as
\begin{equation}
D_i\psi_q(\zbf x)\equiv \left(-i\partial_i-e_qA_i(\zbf x)\right)\psi_q
(\zbf x).
\end{equation}
We shall find it convenient to write the above in the Fourier transform space.
With
\begin{equation}
A_i(\zbf x)=\frac{1}{(2\pi)^{3/2}}\int\tilde A_i(\zbf k)e^{i\zbf k\cdot
\zbf x}d\zbf k,
\end{equation}
we can write the Fourier transform of covariant derivative as
\begin{equation}
\frac{1}{(2\pi)^{3/2}}\int D_i\psi_q(\zbf x)\exp^{-i\zbf k\cdot\zbf x}
d\zbf x=\int d\zbf k'\big[\zbf k_i\delta(\zbf k-\zbf k')
-\frac{e_q}{(2\pi)^{3/2}}\tilde A_i(\zbf k-\zbf k')\big]\tilde\psi_q(\zbf k').
\end{equation}
We now define the continuous matrix
\begin{equation}
\zbf K_i(\zbf k,\zbf k')\equiv\zbf k_i\delta(\zbf k-\zbf k')
-\frac{e_q}{(2\pi)^{3/2}}\tilde A_i(\zbf k-\zbf k').
\label{kmtr}
\end{equation}
The usual covariant derivative for minimal coupling in momentum space
thus consists of the substitution
\begin{equation}
\zbf k_i\tilde\psi_q(\zbf k)\rightarrow\int\zbf K_i(\zbf k,\zbf k')
\tilde\psi_q(\zbf k')d\zbf k'.
\label{cvpsi}
\end{equation}
Let us however note that
\begin{equation}
\tilde\psi_q(\zbf k)=U(\zbf k)q_I(\zbf k)+V(-\zbf k)\tilde q_I(-\zbf k).
\end{equation}
Let us generalise the above with minimal coupling for {\it two component}
operators $q_I(\zbf k)$ and $\tilde q_I(-\zbf k)$ as proposed earlier
\cite{spm78}. Then the
corresponding Fourier transform of the Dirac field is given as
\begin{equation}
\tilde\psi_q^{minml}(\zbf k)=U(\zbf K)q_I(\zbf k)+V(-\zbf K)\tilde q_I(-\zbf
k),
\label{minml}
\end{equation}
where the superscript $minml$ stands for including two component  minimal
electromagnetic coupling, and, $\zbf K$ is a matrix as in equation
(\ref{kmtr}). The difference between equation (\ref{cvpsi}) and (\ref{minml})
is that in the later case the spinors themselves shall contain the covariant
derivatives.

We shall now show that the above result follows due to vacuum realignment when
$|vac'\rangle$ remains gauge invariant. To keep $|vac'\rangle$ gauge
invariant, we replace equation (\ref{b0}) by
\begin{equation}
B_0 ^\dagger=\int q^{0i}_I(\zbf k)^\dagger (h_i(\bfm\sigma\cdot\zbf K))(\zbf k,
\zbf k')\tilde q^{0i}_I(-\zbf k')d\zbf k d\zbf k'.
\label{b1}
\end{equation}
In the above we have replaced $\zbf k^2$ by the matrix $\big(\bfm\sigma
\cdot\zbf K\big)^2$ {\it everywhere} for minimal coupling, since in Dirac
equation the square of momentum always arises through $(\bfm\sigma\cdot
\zbf k)^2=\zbf k^2$, and, have replaced $h_i(\zbf k)(\bfm\sigma\cdot\hat
\zbf k)$ by the matrix $h_i(\bfm\sigma\cdot\zbf K)$. Clearly equation
(\ref{b1}) maintains
gauge invariance of $|vac'\rangle$ at two component level, and, for massless
quarks, gauge invariance at two compenent level and at usual four component
level are equivalent since here the spinors are constants. Now the Bogoliubov
transformations (\ref{bgt}) contain covariant derivatives, and
thus in $|vac'\rangle$ basis the spinors of equation (\ref{uv}) with the
notations of equation (\ref{fqgq}) are given as
\begin{eqnarray}
U(\zbf K )=&&\left(\begin{array}{c} f(\bfm\sigma\cdot\zbf K) \\
(\bfm\sigma \cdot\zbf K) g(\bfm\sigma\cdot\zbf K)\end{array}\right),
\nonumber\\ \nonumber\\
V(-\zbf K )=&&\left(\begin{array}{c} -(\bfm \sigma \cdot \zbf K)
g(\bfm\sigma\cdot\zbf K) \\ f(\bfm\sigma\cdot\zbf K)\end{array}\right).
\label{uv1}
\end{eqnarray}
This shows that the two component minimal coupling of equation
(\ref{minml}) gets generated when $|vac'\rangle$ is gauge invariant.
Also from equation (\ref{sin2h}) we have here
\begin{equation}
sin(2h_i(\bfm\sigma\cdot\zbf K))=cos(\chi_i(\bfm\sigma\cdot\zbf K))
=e^{-\hf R_i^2(\bfm\sigma\cdot\zbf K)^2},
\label{sin2h1}
\end{equation}
which will generate additional gauge couplings of quarks to photons.

The $\pi^0$ state of zero momentum is given as
\begin{equation}
|\pi^0(\zbf 0)\rangle=\frac{1}{2\sqrt 3}\int q_I(\zbf k)^\dagger\tau_3
\tilde q_I(-\zbf k)\tilde u_\pi(\zbf k)d\zbf k|vac'\rangle.
\end{equation}
In the above, $q_I^\dagger$ and $\tilde q_I$ are the two component isospin
(u,d) doublet quark and antiquark creation operators with the spin,
isospin and colour indices having been suppressed.
Thus with minimal coupling the part of the Hamiltonian that gives rise to
the process $\pi^0\rightarrow 2\gamma$ is given as
\begin{equation}
H^{min}=\int\tilde q_I(-\zbf k')^\dagger \big(V(-\zbf K)^\dagger
(\bfm\alpha\cdot\zbf K)U(\zbf K)\big)
(\zbf k',\zbf k) q_I(\zbf k)d\zbf k d\zbf k'.
\label{hmin}
\end{equation}
We then obtain in the lowest order the S-matrix element for the
corresponding transition as
\begin{eqnarray}
\langle\zbf k_1,&&\lambda_1,\zbf k_2,\lambda_2|S|\pi^0(\zbf 0)\rangle
\equiv\delta(P_f-P_i)M_{fi}\nonumber\\
=&&-i\times \frac{1}{2\sqrt 3}\times 2\pi \delta(E_f-E_i)\times 3\times
\nonumber\\
&&\int \langle\zbf k_1,\lambda_1,\zbf k_2,\lambda_2|
\sum_{q=u,d}Tr\Big(V(-\zbf K_q)^\dagger(\bfm\alpha\cdot\zbf K_q)\tau_3
U(\zbf K_q)\Big)(\zbf k,\zbf k)|vac'\rangle\tilde u_\pi(\zbf k)d\zbf k.
\label{mfi0}
\end{eqnarray}
We now note that from equation (\ref{uv1}) we have the matrix equation,
\begin{eqnarray}
&& V(-\zbf K)^\dagger(\bfm\alpha\cdot\zbf K) U(\zbf K)\nonumber\\
=&& f(\bfm\sigma\cdot\zbf K)(\bfm\sigma\cdot\zbf K) f(\bfm\sigma\cdot\zbf K)
-g(\bfm\sigma\cdot\zbf K)(\bfm\sigma\cdot\zbf K)^3
g(\bfm\sigma\cdot\zbf K)\nonumber\\
=&& \Big(2f(\bfm\sigma\cdot\zbf K)^2-1\Big)(\bfm\sigma\cdot\zbf K)
=e^{-\hf R^2(\bfm\sigma\cdot\zbf K)^2}(\bfm\sigma\cdot\zbf K).
\label{vdgu}
\end{eqnarray}
In the last equation we have used \cite{spm78} the identity
$f(x)^2+x^2g(x)^2=1$.
We note the special form of the above equation as being directly related
to the vacuum structure through equation (\ref{sin2h}) or (\ref{sin2h1}).

Clearly in equation (\ref{vdgu}) we have {\it any} number of photon fields,
and, we are to
choose only two of them to obtain the matrix element as in equation
(\ref{mfi0}). We shall do so using that
\[\langle\zbf k_1,\lambda_1|\tilde A_i(\zbf k-\zbf k')|vac'\rangle
=\frac{1}{\sqrt {2|\zbf k_1|}}\delta(\zbf k_1+\zbf k-\zbf k')
e_i(\zbf k_1,\lambda_1).\]
For example, we first note that
\begin{eqnarray}
&&\langle \zbf k_1,\lambda_1,\zbf k_2,\lambda_2|Tr\big[\big(\bfm\sigma\cdot
\zbf K_q\big)^3(\zbf k,\zbf k)\big]|vac'\rangle \nonumber\\
=&&\delta(\zbf k_1+\zbf k_2)\times 2i\epsilon_{imj}\cdot\frac{e_q^2}
{(2\pi)^3}\cdot\frac{(\zbf k^m-\zbf k_2^m)-(\zbf k^m-\zbf k_1^m)}
{2|\zbf k_1|}\times e_i(\zbf k_1,\lambda_1)e_j(\zbf k_2,\lambda_2)
\times\frac{1}{\sqrt 2},
\end{eqnarray}
and that
\begin{eqnarray}
&&\langle \zbf k_1,\lambda_1,\zbf k_2,\lambda_2|Tr\big[\big(\bfm\sigma\cdot
\zbf K_q\big)^5(\zbf k,\zbf k)\big]|vac'\rangle \nonumber\\
=&&\delta(\zbf k_1+\zbf k_2)\times 2i\epsilon_{imj}\cdot\frac{e_q^2}
{(2\pi)^3}\cdot\frac{2\zbf k_1^m}
{2|\zbf k_1|}\times e_i(\zbf k_1,\lambda_1)e_j(\zbf k_2,\lambda_2)
\times\frac{1}{\sqrt 2}.(3k^2+q^2),
\end{eqnarray}
where, $k^2=\zbf k^2$, and, $q^2=(\zbf k-\zbf k_1)^2$, and we
are using that $\tilde u_\pi(\zbf k)$ is even in $\zbf k$. Proceeding
in a similar manner, from equation (\ref{mfi0}) we then obtain that
\begin{equation}
M_{fi}=\frac{\alpha}{\pi}\cdot\frac{1}{2\sqrt 3}\cdot\epsilon_{imj}\cdot
\frac{\zbf k_1^m}{|\zbf k|}e_i(\zbf k_1,\lambda_1)e_j(\zbf k_2,\lambda_2)
\cdot R^2I(R,m_\pi),
\label{mfi1}
\end{equation}
where $I$ is given as, with $|\zbf k_1|=m_\pi/2$,
\begin{eqnarray}
I(R,m_\pi)=&&\int \tilde u_\pi(\zbf k)d\zbf k\Big[1
-\frac{1}{2!}\left(\frac{R^2} {2}\right)(3k^2+q^2)
+\frac{1}{3!}\left(\frac{R^2} {2}\right)^2(5k^4+3k^2q^2+q^4)\nonumber\\
&&-\frac{1}{4!}\left(\frac{R^2} {2}\right)^3(7k^6+5k^4q^2+3k^2q^4+q^6)
\nonumber\\
&&+\frac{1}{5!}\left(\frac{R^2} {2}\right)^4(9k^8+7k^6q^2+5k^4q^4+3k^2q^6
+q^8)+\cdots\Big].
\label{int}
\end{eqnarray}
In the above we have used that $3(e_u^2-e_d^2)=e^2$ as indicating separately
the contributions from $u$ and $d$ quarks. We know the wave function
$\tilde u_\pi(\zbf k)$, and hence calculating the above integral
we obtain that $I=0.0858$ GeV$^{3/2}$. The decay width then becomes
\begin{equation}
\Gamma(\pi^0\rightarrow 2\gamma)=\frac{m_\pi^2}{4}\cdot\sum_{\lambda_1
\lambda_2}\left|M_{fi}\right|^2=m_\pi^2\cdot\frac{\alpha^2}{48\pi^2}
\cdot R^4I^2\approx 8.3\;{\rm eV}.
\label{pi0gg}
\end{equation}
This may be compared with the experimental value given as
7.7$\pm$0.6 eV. The agreement here {\it without} fixing any parameter thus
appears to be rather good. The anomaly equation has not been used, and
the vacuum structure as determined through $f_\pi$ seems to be adequate.

It thus appears that the two component minimal coupling contains within itself
the effects of anomaly! Instead of the anomaly equation,
gauge invariance while considering restructuring of vacuum
does the job. This may have consequences beyond
the verification of the above width for the $\pi^0$ decay. Here as noted
gauge invariance at the two component level {\it gets generated}
since the spinors are related to the vacuum structure. The equation
(\ref{pi0gg}) in the context of chiral symmetry restoration in quark gluon
plasma is further discussed in section {\bf VII}.

\section{ Masses of mesons}
In the last section we have taken  the wave function of the pion as
given through chiral symmetry breaking. Simultaneously, in the
expressions involving the pion mass, we have naturally substituted the
mass of the pion as observed. However, we know that the pion is only
{\it approximately} a Goldstone boson, and that $m_u,m_d$
do not vanish. We shall now include this effect to relate
Lagrangian masses with the masses of the mesons. We shall however
avoid the diagonal pseudoscalar mesons, where anomaly effects from
the gluonic contributions will be there \cite{donogh}, which are not
being considered in the present paper.

We now take the mass part of the Hamiltonian density as
\begin{equation}
{\cal H}_{mass}(\zbf x)=\sum m_i\bar\psi_i(\zbf x)\psi_i(\zbf x).
\label{hmass}
\end{equation}
As before, $i$ is the flavour, and summation over colour is understood.
 From the definition of the pion state as obtained through chiral symmetry
breaking in equation (\ref{pipst}), we then obtain for example the mass
of $\pi^+$ as, using translational invariance,
\begin{eqnarray}
m(\pi^+)=&& (2\pi)^3\cdot\frac{N_\pi^2}{6}\langle vac'|Q_{\pi^+}^\dagger
{\cal H}_{mass}(\zbf x)Q_{\pi^+}|vac'\rangle \nonumber\\
\equiv && (2\pi)^3\cdot\frac{N_\pi^2}{6}\cdot\hf\cdot\langle vac'|\big[\big[
Q_{\pi^+}^\dagger,{\cal H}_{mass}(\zbf x)\big],Q_{\pi^+}\big]|vac'\rangle
\nonumber\\
= && (2\pi)^3\cdot\frac{N_\pi^2}{6}\cdot\hf\cdot(m_1+m_2)\langle vac'|\big[-
\bar\psi_1\psi_1-\bar\psi_2\psi_2\big]|vac'\rangle.
\label{pims0}
\end{eqnarray}
 From the above equation, using the earlier expression for $f_\pi$ as in
equation (\ref{fpi}) we can obtain the conventional current algebra result
that
\begin{equation}
m_\pi^2=\frac{m_u+m_d}{2}\cdot\frac{-\langle\bar uu\rangle-\langle\bar dd
\rangle}{f_\pi^2}.
\label{mfpi}
\end{equation}
However, with the present ansatz we also have a simpler relationship when
$R_1=R_2$ with the pion mass expressed in terms of the current quark
masses. This is given from equations (\ref{qcnd}) and (\ref{pims0}) as
\begin{equation}
m_\pi=\frac{m_u+m_d}{2}\cdot 4\sqrt 2,
\label{pims1}
\end{equation}
where, the explicit dependance on $R$ cancels out. Equation
(\ref{pims1}) yields that
\begin{equation}
\frac{m_u+m_d}{2}\approx 24\; {\rm MeV}.
\label{udm}
\end{equation}
We note that the current quark masses for $u$ and $d$ quarks
are rather large, and, here the pion mass becomes
independent of the condensate scale.

In the same way we can also get the current quark mass $m_s$. For this
purpose we note that, with $N_K$ given by equation (\ref{normk}),
\begin{eqnarray}
m(K^+)=&& (2\pi)^3\cdot\frac{N_K^2}{6}\cdot\hf(m_1+m_3)\langle vac'|\big[-
\bar\psi_1\psi_1-\bar\psi_3\psi_3\big]|vac'\rangle \nonumber\\
=&& (2\pi)^3\cdot\frac{N_K^2}{6}\cdot\hf(m_1+m_3)
\times\frac{3}{\sqrt 2\pi{\sqrt\pi}}\times\left(\frac{1}{R_1^3}+\frac{1}{R_3^3}
\right).
\label{kms0}
\end{eqnarray}
This yields that
\begin{equation}
m_3=m_s\approx 96 \;{\rm MeV}.
\label{sm}
\end{equation}
The current quark mass for the strange quark here appears to be smaller
than usual. From equation (\ref{fk})
we may also obtain the conventional current algebra result that
\begin{equation}
m_K^2=\frac{m_u+m_s}{2}\cdot\frac{-\langle\bar uu\rangle-\langle\bar ss
\rangle}{f_K^2}.
\label{mfk}
\end{equation}

We may compare the {\it explicit} symmetry breaking parameters $m_i$ with the
vacuum realignment scales $1/R_i$. We then note that for example
\begin{equation}
\frac{m_u+m_d}{2}\cdot R_1=0.118,\qquad {\rm and}\qquad
m_s\cdot R_s=.194.
\end{equation}
Hence when we determine the wave function from equations like (\ref{qivcp})
with breaking of exact chiral symmetry, we expect the $K$-meson wave
function to be less reliable than for the pion. As a curiousity, {\it if}
there were a pseudoscalar meson with only strange quarks, parallel to equation
(\ref{pims1}) we shall have
\begin{equation}
m(\bar ss)=m_s\times 4\sqrt 2=543\;{\rm MeV}.
\label{mss}
\end{equation}
We note that the isospin diagonal mass matrix shall in addition have gluonic
anomaly contributions \cite{donogh} and equation (\ref{mss}) does not
include the same. There will really be a mass matrix connecting to other
states. We shall not deal with the same here.

We will however still adopt the same philosophy for the heavier
nondiagonal mesons as a `zeroeth order' approximation.
For these mesons, the corresponding decay constants are not known.
Hence we adopt a different strategy. We note that for each quark here
we are introducing only two parameters.  We first introduce an assumed
vacuum structure given by the parameter $R_i$, and then introduce
the Lagrangian quark masses $m_i$ to obtain the finite masses of the
mesons. For the D-mesons we know that $m(D^+)=1869$ MeV, and that
$m(D_s)=$1968 MeV. Let us now try to determine the values of
$R_4=R_c$ and $m_4=m_c$. We note that as earlier the $D^+$ state is given
as
\begin{eqnarray}
\mid D^+(\zbf 0)>&&=N_{D}\cdot\frac{1}{\sqrt6}\int c_I(\zbf k)^\dagger
{\tilde d_I}(-\zbf k)\big[cos(\hf\chi_2(\zbf k))cos(\hf\chi_4(\zbf k))
\nonumber\\ &&
-sin(\hf\chi_2(\zbf k))sin(\hf\chi_4(\zbf k))\big]
d\zbf k \mid vac'>,
\label{dpst}
\end{eqnarray}
where the normalisation constant $N_D$ can be calculated.
The $D$-meson mass is now given as
\begin{eqnarray}
m(D^+)=&& (2\pi)^3\cdot\frac{N_D^2}{6}\cdot\hf(m_2+m_4)\langle vac'|\big[-
\bar\psi_2\psi_2-\bar\psi_4\psi_4\big]|vac'\rangle \nonumber\\
=&& \frac{N_D^2}{6}\cdot\hf(m_d+m_c)
\times\frac{3\pi^{3/2}}{\sqrt 2}\times\left(\frac{1}{R_2^3}+\frac{1}{R_4^3}
\right).
\label{dms0}
\end{eqnarray}
Similarly, the $D_s$ mass is given as
\begin{equation}
m(D_s^+)=\frac{N_{D_s}^2}{6}\cdot\hf(m_s+m_c)
\times\frac{3\pi^{3/2}}{\sqrt 2}\times\left(\frac{1}{R_s^3}+\frac{1}{R_c^3}
\right).
\label{dsms0}
\end{equation}
We note that parallel to equation (\ref{normk}) the normalisations
$N_D$ and $N_{D_s}$ are given through
\begin{equation}
\frac{1}{N_D^2}=2\pi\int\Big[1-sin(\chi_2(|\zbf k|))sin(\chi_4(|\zbf k|))
+cos(\chi_2(|\zbf k|))cos(\chi_4(|\zbf k|))\Big]|\zbf k|^2d|\zbf k|,
\label{normd}
\end{equation}
and,
\begin{equation}
\frac{1}{N_{D_s}^2}=2\pi\int\Big[1-sin(\chi_3(|\zbf k|))sin(\chi_4(|\zbf k|))
+cos(\chi_3(|\zbf k|))cos(\chi_4(|\zbf k|))\Big]|\zbf k|^2d|\zbf k|.
\label{normds}
\end{equation}
We now use equations (\ref{dms0}) and (\ref{dsms0}) along with the appropriate
masses for the particles to solve for the new parameters $m_c$ and $R_c$
numerically. We then obtain that
\begin{equation}
m_4=m_c\approx 355\;{\rm MeV}; \qquad R_4^2=R_c^2=0.4852\;{\rm GeV}^{-2}.
\label{mcrc}
\end{equation}
We also note that here
\begin{equation}
m_c\cdot R_c=0.247.
\end{equation}
As stated earlier,
this can be regarded as a measure of error in the wave function obtained
from exact chiral symmetry breaking.
We also predict the corresponding decay constants to be
\begin{equation}
f_D=248\; {\rm MeV}; \qquad f_{D_s}=261\;{\rm MeV}.
\label{fdds}
\end{equation}

We use the same strategy for the $B$-mesons. We note that
$m_B=5.277$ GeV, and that $m_{B_s}=5.383$ GeV. As earlier
we take $m_b$ and $R_b$ as unknown, and then determine the same from
the known values of the above masses. This yields that
\begin{equation}
m_b\approx 1.04\;{\rm GeV}; \qquad R_b^2=0.2307\;{\rm GeV}^{-2}.
\label{mbrb}
\end{equation}
Here $m_bR_b=0.50$, which is rather large. We also obtain that
$m_{B_c}=5.586$ GeV, and, the corresponding decay constants as
\begin{equation}
f_B=256\;{\rm MeV}; \qquad f_{B_s}=261\;{\rm MeV}; \qquad
f_{B_c}=475\;{\rm MeV}.
\label{fbbs}
\end{equation}

We note that in the above as the current quark mass increases, the
product $m_iR_i$ becomes larger and larger. Hence to describe the meson
wave functions through only the chiral symmetry breaking mechanism
becomes less and less valid. For heavy mesons, it is only a `zeroeth order'
approximation, but the surprising feature is that always the masses
from chiral symmetry breaking dominate over the current quark masses.
Hence with the present hypothesis the heavy quarks can not be
approximated by Dirac spinors. This shall have an effect on
spectroscopic properties for the spin dependance of
the energy levels for heavy hadrons.

\section{ Pion and kaon charge radii}
We shall here briefly recall the calculation of the pion charge radius
\cite{amspm} and then obtain the same for the kaon. The pion form
factor is given by the equation
\cite{spm78}
\begin{equation}
<\pi^+(-\zbf p)|J^0(0)|\pi^+(\zbf p)>=\frac{1}{(2\pi)^3}\cdot \frac{m_{\pi}}
{p^0}\cdot G^{\pi}_E(t)
\label{formftr}
\end{equation}
where through direct evaluation \cite{spm78} we obtained that
\begin{equation}
G_E^{\pi}(t)=\sum_q e_q\int \tilde u_{\pi}(\zbf k'_1)^\dagger
\tilde u_{\pi}(\zbf k_1)
\left(f_q(\zbf k_1')f_q(\zbf k_1)+\zbf k_1'\cdot\zbf k_1g_q(\zbf k_1')
g_q(\zbf k_1)\right)d\zbf k_1.
\label{piform}
\end{equation}
In the above, $q$ stands for $u$ or $d$ quark, and, \cite{spm78},
\begin{equation}
t=-4\zbf p^2; \qquad \zbf k_1'=\zbf k_1-\frac{m_{\pi}}{p^0}\zbf p.
\end{equation}
For the charge radius we retain terms only upto $|\zbf p|^2$ and identify
the same through the equation
\begin{equation}
G_E(t)=1+\frac{R_{ch}^2}{6}\;t\;.
\label{gchr}
\end{equation}
We now substitute $\zbf k_1'=\zbf k -\frac{1}{2}\zbf p$ and
$\zbf k_1=\zbf k +\frac{1}{2}\zbf p$. Upto order $|\zbf p|^2$, along with
other equations we also have \cite{amspm}
\begin{equation}
\hat k_1'\cdot\hat k_1=1-\frac{\zbf p^2}{3\zbf k^2}.
\label{k1pk1}
\end{equation}
On simplification, we had finally obtained \cite{amspm}
\begin{equation}
R_{ch}^2=R_{ch1}^2+R_{ch2}^2
\label{rch12}
\end{equation}
where,
\begin{equation}
R_{ch1}^2=\frac{N_{\pi}^2}{4}\int (\zbf \bigtriangledown cos\chi(\zbf k))^2
d\zbf k
\label{rch1}
\end{equation}
is the contribution coming from the wave function alone, and,
\begin{equation}
R_{ch2}^2=\frac{N_{\pi}^2}{16}\int cos^2\chi(\zbf k)
\left[\frac{R_1^4\zbf k^2cos^2 \chi(\zbf k)}{1-cos^2\chi(\zbf k)}
+\frac{4(1-cos\chi(\zbf k))}{\zbf k^2}\right]d\zbf k
\label{rch2}
\end{equation}
is the balance of the contribution. In the second term of the right hand
side above we have corrected \cite{amspm} a factor two.

For the choice of equation (\ref{piwvf}) and the expression for
$cos\chi(\zbf k)=cos\chi_1(\zbf k)$, we then obtain that $R_{ch1}^2
= 8.842$ GeV$^{-2}$, and $R_{ch2}^2= 2.591$ GeV$^{-2}$. We thus obtain,
\begin{equation}
R_{ch}=0.67\;{\rm fms},
\end{equation}
which may be compared with the experimental value of $R_{ch}=0.66$ fms
\cite{rchpi}. The result is quite good as compared to the last calculation
\cite{amspm} on correcting the error.

For the $K$-meson, finding out the charge radius is more
complicated. The form factor to be considered here is given through
\begin{equation}
<K^+(-\zbf p)|J^0(0)|K^+(\zbf p)>=\frac{1}{(2\pi)^3}\frac{m_{K}}
{p^0}G^K_E(t)
\label{formftk}
\end{equation}
Parallel to equation (\ref{piform}), with
$q$ quark as the interacting quark, we have
\begin{equation}
G_E^K(t)= \sum_q e_q\int \tilde u_K(\zbf k'_1)^\dagger
\tilde u_K(\zbf k_1)
\left(f_q(\zbf k_1')f_q(\zbf k_1)+\zbf k_1'\cdot\zbf k_1g_q(\zbf k_1')
g_q(\zbf k_1)\right)d\zbf k_1,
\label{kform}
\end{equation}
for $u$ and $s$ quark with momenta as below.
The state of finite momentum is constructed here with Lorentz
boosting \cite{spm78}, where we need the energy shared by the quark and
antiquark at rest to obtain the time dependance of the operators for
boosting. A specific threshold enhancement arising from this feature
has been noted earlier \cite{spmthr}.
In the pion both quark and antiquark equally share the pion rest
energy which gives the energy shared as half the pion mass.
However, kaon being an asymmetric system, the energy shared by the quarks
shall be different, and the time dependance is unknown \cite{spm78}.
This gives an ambiguity for the determination of the charge radius.

Let $k_1$ and $k_2$ be the
{\it four-momenta} of the $u$ and $\bar s$ for $K^+$ at rest \cite{spm78}
where $k_1^0=\lambda_1m_K$ and $k_2^0=\lambda_2m_K$ with $\lambda_1+
\lambda_2=1$. For the
state $|K^+(\zbf p)\rangle$, the above momenta will be transformed through
the Lorentz matrix $L(p)$ taken as, with $\mu=0,\cdots,3$ and
$i,j=1,2,3$ \cite{spm78}
\begin{equation}
L(p)_{\mu 0}=L(p)_{0\mu }=\frac{p^\mu}{m_K}\quad {\rm and}\quad
L(p)_{ij}=\left(\delta_{ij}+\frac{p^ip^j}{m_K(p^0+m_K)}\right).
\label{lpk}
\end{equation}
In the above $p^0=\left(\zbf p^2+m_K^2\right)^{1/2}$. Similarly for
the state $|K^+(-\zbf p)\rangle$ let the quark momenta at rest
be $k_1'$ and $k_2'$ which are Lorentz boosted by matrix $L(p')$ with
$\zbf p'=-\zbf p$. We note that we have also
$k_1'^0=\lambda_1m_K$ and $k_2'^0=\lambda_2m_K$.
Let $G_E^{K1}(t)$ be the contribution for form factor where $u$-quark
interacts. Since here $\bar s$ is spectator, we get momentum conservation
equation as \cite{spm78}
\begin{equation}
L(p)_{ij}k_2^j+L(p)_{i0}k_2^0=L(p')_{ij}k_2'^j+L(p')_{i0}k_2^0 .
\end{equation}
On multiplyng the above by the inverse of the $3\times3$ matrix
$L(p)_{ij}=L(p')_{ij}$, we then obtain that
\begin{equation}
\zbf k_2'=\zbf k_2+\lambda_2\frac{m_K}{p^0}\cdot 2\zbf p, \qquad{\rm or,}
\qquad\zbf k_1'=\zbf k_1-\lambda_2\frac{m_K}{p^0}\cdot 2\zbf p.
\label{k1k1p}
\end{equation}
We then replace $\zbf k_1$ by the symmetric integration variable
$\zbf k$ with the substitutions
\begin{equation}
\zbf k_1'=\zbf k-\lambda_2\frac{m_K}{p^0}\zbf p;
\qquad
\zbf k_1=\zbf k+\lambda_2\frac{m_K}{p^0}\zbf p.
\label{ksym}
\end{equation}
Thus, when the $u$-quark interacts we obtain that
\begin{equation}
G_E^{K1}(t)= e_u \int \tilde u_K(\zbf k'_1)^\dagger \tilde u_K(\zbf k_1)
\left(f_1(\zbf k_1')f_1(\zbf k_1)+\zbf k_1'\cdot\zbf k_1g_1(\zbf k_1')
g_1(\zbf k_1)\right)d\zbf k,
\label{kform1}
\end{equation}
where the variables are as given in equation (\ref{ksym}).
We also note that spin rotations have been included \cite{spm78}, and,
there is no contribution from the same as we have here $S(L(p'))^\dagger
S(L(p))=I$ for $\zbf p'=-\zbf p$.

There will be a parallel contribution $G_E^{K2}(t)$ where the $\bar s$
interacts and $u$ is the spectator. This contribution is obtained in a
similar manner as
\begin{equation}
G_E^{K2}(t)=- e_s \int \tilde u_K(\zbf k'_1)^\dagger \tilde u_K(\zbf k_1)
\left(f_3(\zbf k_1')f_3(\zbf k_1)+\zbf k_1'\cdot\zbf k_1g_3(\zbf k_1')
g_3(\zbf k_1)\right)d\zbf k,
\label{kform2}
\end{equation}
where $\lambda_2\rightarrow\lambda_1$, and $e_u\rightarrow -e_s$.
Here parallel to equation (\ref{ksym}) we have
$\zbf k_1'=\zbf k-\lambda_1\frac{m_K}{p^0}\cdot \zbf p$, and,
$\zbf k_1=\zbf k+\lambda_1\frac{m_K}{p^0}\cdot \zbf p$.
On simplification we then obtain that
\begin{equation}
R_{chK}^2=R_{chK1}^2+R_{chK2}^2
\label{rchk12}
\end{equation}
where, parallel to equation (\ref{rch1})
\begin{equation}
R_{chK1}^2=\left(\frac{2}{3}\lambda_2^2+\frac{1}{3}\lambda_1^2\right)\times
\int (\zbf \bigtriangledown \tilde u_K(\zbf k))^2 d\zbf k
\label{rchk1}
\end{equation}
is the contribution coming from the wave function of equation (\ref{kwvf}),
and,
\begin{eqnarray}
R_{chK2}^2=&&\frac{2}{3}\times\lambda_2^2\int u_K(\zbf k)^2
\left\{\frac{R_1^4\zbf k^2cos^2(\chi_1(\zbf k))}{4(1-cos^2\chi_1(\zbf k))}
+\frac{(1-cos\chi_1(\zbf k))}{\zbf k^2}\right\}d\zbf k \nonumber\\
\nonumber\\
+&&\frac{1}{3}\times\lambda_1^2\int u_K(\zbf k)^2
\left\{\frac{R_3^4\zbf k^2cos^2(\chi_3(\zbf k))}{4(1-cos^2\chi_3(\zbf k))}
+\frac{(1-cos\chi_3(\zbf k))}{\zbf k^2}\right\}d\zbf k.
\label{rchk2}
\end{eqnarray}
is the balance of the contribution.
We may easily note that when $R_1=R_3$ and $\lambda_1=\lambda_2=\hf$, the
above expressions go over to the corresponding expressions for the pion.
The first term in the curly brackets above came from the simplification
\begin{eqnarray}
&& cos\left(\hf\chi_1(\zbf k_1')\right)cos\left(\hf\chi_1(\zbf k_1)\right)
+sin\left(\hf\chi_1(\zbf k_1')\right)sin\left(\hf\chi_1(\zbf k_1)\right)
\nonumber\\
\approx &&  1-\frac{R_1^4\zbf k^2cos^2\chi_1(\zbf k)}
{6(1-cos^2\chi_1(\zbf k))}\times\lambda_2^2\zbf p^2,
\end{eqnarray}
and, the second term came from
\begin{equation}
sin^2\left(\hf\chi_1(\zbf k)\right)\hat\zbf k_1'\cdot\hat\zbf k_1
\rightarrow -\frac{2(1-cos\chi_1(\zbf k))}{3\zbf k^2}\times
\lambda_2^2\zbf p^2.
\end{equation}
The above contributions are for $\bar s$ being the spectator,
and, we have used equations (\ref{sin2h}) and (\ref{ksym}) for the
simplification. The second curly bracket arises on
interchanging the two quarks.

We now have to estimate $\lambda_1$ and $\lambda_2$.
We had earlier suggested \cite{sp86} that for sharing
of the energy at rest, the kinetic energies of the two constituents
may be different, but the potential energy shall be equally shared.
In the present determination of the mass of the kaon the potential picture
is absent. We shall however extrapolate the same by looking at the
expression in (\ref{kms0}) to guess these factors. We shall
consider here two possible identifications. From equation (\ref{kms0}) let
us ``identify" the potential energy as
\begin{equation}
v_K=(2\pi)^3\cdot\frac{N_K^2}{6}\cdot\hf\big[
m_1\langle vac'|- \bar\psi_3\psi_3|vac'\rangle
+m_3\langle vac'|- \bar\psi_1\psi_1|vac'\rangle\big].
\end{equation}
The balance of the contributions contain only $u$ terms or only $s$ terms,
which we identify as the respective kinetic contributions. We then obtain that
\cite{sp86} $\lambda_1=0.134$ and $\lambda_2=0.866$. This yields that
$R_{chK1}^2=4.39$ GeV$^{-2}$ and $R_{chK2}^2=1.20$ GeV$^{-2}$, so that
\begin{equation}
R_{chK}=0.47\;{\rm fms.}
\label{chk1}
\end{equation}
We may otherwise identify that in equation (\ref{kms0}) the $\langle \bar uu
\rangle$ part corresponds to $\lambda_1 m_K$, and, the
$\langle \bar s s\rangle$ part corresponds to $\lambda_2 m_K$. We then have
$\lambda_1=0.067$ and $\lambda_2=0.933$. This yields that
$R_{chK1}^2=5.04$ GeV$^{-2}$ and $R_{chK2}^2=1.37$ GeV$^{-2}$, so that
\begin{equation}
R_{chK}=0.50\;{\rm fms.}
\label{chk2}
\end{equation}
The above values may be compared with the experimental value of $R_{chK}=0.58$
fms \cite{donogh}. The calculated value appears to be small, and indicates that
taking the wave function as determined from {\it exact}
chiral symmetry breaking may not be correct. The ``identification"
of the fractions $\lambda_1,\lambda_2$ is also unreliable. We
{\it should} have a handle on spectroscopy which may clarify the above
as well as give corrections to the kaon wave function as different from
the purely vacuum structure contribution as in equation (\ref{kwvf}).

\section{discussions}
Let us recall what has been achieved here. We first neglect masses of the
quarks and assume that global chiral symmetry breaks spontaneously. This
is described through a vacuum realignment where we approximate
for the correlation of a quark at two different space points as in
equations (\ref{expqq}) and (\ref{sin2h}) with a Gaussian function.
Thus for the vacuum structure for quark $q_i$, we introduce a single
parameter $R_i$. This parameter also gives the four component quark
field operators for exact chiral symmetry breaking.
We then find that $R_i$ can be determined from the experimental value
of the decay constant. Further, for approximate chiral
symmetry breaking, we obtain the
masses of the mesons through current algebra as in equations
like (\ref{kms0}) or (\ref{mfk}) in terms of the Lagrangian masses of the
quarks. This introduces another constant $m_i$ for each quark. With only
these two parameters for each quark, we use the explicit form of vacuum
realignment to draw conclusions for the hadronic properties and examine
consistency of such a hypothesis.

For the light quark sector, we find that the parameter $R$ for the vacuum
structure as determined from $f_\pi$ also yields $\Gamma(\pi^0
\rightarrow 2\gamma)$ and $R_{ch}^2$ correctly. For the $\pi^0$ decay,
we use the fact that the destabilised vacuum $|vac'\rangle$ should
remain gauge invariant. For the charge radius, we use that mesons in motion
should be obtained through Lorentz boosting
\cite{spm78}. It thus appears that {\it we know} the vacuum structure
for quark condensates of the light quark sector from experimental
observations of the above hadronic properties as conjectured earlier
\cite{amspm}.

We also determine the vacuum structure of the $s$-quark sector from
the experimental value of $f_K$, and, using the same, go on to derive
the charge radius of the kaon. This falls short of the experimental value
by about fifteen percent. We may recall that from chiral perturbation
theory a similar disagreement is also there, where the theoretical value
is larger by about the same amount \cite{donogh}.

We have next
calculated the decay constants of $D$ and $B$ mesons. In this sector a
quantitative agreement of the same is not expected.
We find however that with the present hypothesis, the spinor
structure for $c$ and $b$ quarks may be quite different from that of a free
Dirac particle. This feature shall have consequences for spectroscopy of
heavy mesons.

Let us now note some obvious limitations of the present calculations. The
$s$-quark seems to have a Lagrangian mass of the order of 100 MeV, and
the corresponding
masses of $c$ and $b$ quark are higher. For them we have considered the
wave functions of the mesons as obtained totally from chiral symmetry
breaking. In some sense this may not be very bad as {\it post facto} the
scales for chiral symmetry breaking for them are higher than the above
masses. It is however
desirable to look further into this. A related assumption has been that
for all the quarks we have taken $\chi_0(\zbf k)=\frac{\pi}{2}$. On the
other hand, as per equation (\ref{chi0}) we could have taken some thing
like $cos\chi_0(\zbf k)=m_i/\epsilon_0(\zbf k)$ and
$sin\chi_0(\zbf k)=|\zbf k|/\epsilon_0(\zbf k)$ with $\epsilon_0(\zbf k)
=\sqrt{\zbf k^2+m_i^2}$, which corresponds to free fields for
$m_i\ne 0$. We have not done the same here since our main objective has
been to relate the vacuum structure for chiral symmetry breaking to
hadronic properties with simple calculations, and show that this structure
is observable. For light quarks the programme works unexpectedly well.
For heavy quarks, we believe that the refinements mentioned above shall be
inevitable.

As $R_i\rightarrow\infty$, chiral symmetry gets restored. We
note from equation (\ref{pims1}) that the $\pi^0$ mass will remain
unaltered. From equation (\ref{pi0gg}) however with a direct evaluation
we note that then $\Gamma(\pi^0\rightarrow 2\gamma)$ continuously
increases as $R^2$ increases. In particular, with $R^2$ respectively as
30 GeV$^{-2}$, 50 GeV$^{-2}$ and 70 GeV$^{-2}$ for the vacuum structure,
the corresponding values of $\Gamma(\pi^0\rightarrow 2\gamma)$ are
given as 9.2 eV, 11.3 eV and 12.7 eV. This may
be relevant for observing chiral symmetry restoration in quark gluon
plasma (QGP) during an intermediate phase when quarks and gluons and
hadrons are present. Since with temperature $R$ is likely to increase,
this will show up in $\Gamma(\pi^0\rightarrow 2\gamma)$ over its value
at zero temperature and may be looked for as a signal for progress towards
chiral symmetry restoration prior to phase transition.
We note that $J/\psi$ suppression \cite{satz} has been a conventional
signature for QGP \cite{cps}. Can we also calculate and expect
a variation of the $\eta$, $\eta'$ and $\eta_c$ widths? For this
purpose it shall be necessary
to see whether anomaly effects of the gluonic sector \cite{donogh}
can be absorbed through a mechanism similar to what has been
done for the electromagnetic sector for $\pi^0$ decay. The real problem
is to develop a technology parallel of conventional spectroscopy, which
includes in an essential manner approximate chiral symmetry breaking.

Intuitively one might expect that $\pi^0$ dissolves more easily as $R$
increases \cite{satz}. The calculation here seems to illustrate the same.
Also, most of the $\pi^0$ decay will take
place during hadronisation process or later, when the temperature will
be less. From equations (\ref{pinorm}) and (\ref{fpi}) there will also be
a similar dependance of $f_\pi$ on $R^2$.
In contrast to $\pi^0$, as $R^2$ increases $\pi^{\pm}$ progressively
become more stable for the leptonic decay mode.

The present work generalises Ref. \cite{amspm} in being able to obtain
the $\pi^0$ width through a derived form of minimal coupling. It also gives
a theoretical base for the {\it ad hoc} assumptions of Ref. \cite{spm78},
applied to many coherent and incoherent processes \cite{spmcoh,spmincoh}.
It may be desirable to see how these results change with the present form
of equation (\ref{fqgq}) related to vacuum structure of (\ref{sin2h}).
The bright side here is that we now seem to know the vacuum
structure for light quarks. However, it also emphasizes the limitations in
our understanding the same for heavier quarks, while illustrating their
relevance for the corresponding hadronic properties.

\acknowledgements
The author wishes to thank O. Pene, A. N. Kamal, N. Barik, A. R. Panda,
H. Mishra, A. Mishra,
P. K. Panda, S. Mishra and C. Das for many discussions. The author also
acknowledges to the Department of Science and Technology, Government of
India for the project SP/S2/K-45/89.

\def \ltg{ K.G. Wilson, Phys. Rev. \zbf  D10, 2445 (1974); J.B. Kogut,
Rev. Mod. Phys. \zbf  51, 659 (1979); ibid  \zbf 55, 775 (1983);
M. Creutz, Phys. Rev. Lett. 45, 313 (1980); ibid Phys. Rev. D21, 2308
(1980); T. Celik, J. Engels and H. Satz, Phys. Lett. B129, 323 (1983)}

\def \svz {M.A. Shifman, A.I. Vainshtein and V.I. Zakharov,
Nucl. Phys. B147, 385, 448 and 519 (1979);
R.A. Bertlmann, Acta Physica Austriaca 53, 305 (1981)}

\def \spmbst {S.P. Misra, Phys. Rev. D35, 2607 (1987)}

\def \hmgrnv { H. Mishra, S.P. Misra and A. Mishra,
Int. J. Mod. Phys. A3, 2331 (1988)}

\def \snss {A. Mishra, H. Mishra, S.P. Misra
and S.N. Nayak, Phys. Lett 251B, 541 (1990)}

\def \amqcd { A. Mishra, H. Mishra, S.P. Misra and S.N. Nayak,
Pramana (J. of Phys.) 37, 59 (1991); A. Mishra, H. Mishra, S.P. Misra
and S.N. Nayak, Zeit. fur Phys. C 57, 233 (1993); A. Mishra, H. Mishra
and S.P. Misra, Z. Phys. C 58, 405 (1993)}

\def \spmtlk {S.P. Misra, Talk on `Phase transitions in quantum field
theory' in the Symposium on Statistical Mechanics and Quantum field theory,
Calcutta, January, 1992, hep-ph/9212287}

\def \hmnj {H. Mishra and S.P. Misra, Phys. Rev. D 48, 5376 (1993)}

\def \hmqcd {A. Mishra, H. Mishra, V. Sheel, S.P. Misra and P.K. Panda,
hep-ph/9404255 (1994)}

\def \amcrl {A. Mishra, H. Mishra and S.P. Misra, Z. Phys. C 57, 241 (1993)}

\def \higgs { S.P. Misra, in {\it Phenomenology in Standard Model and Beyond},
Proceedings of the Workshop on High Energy Physics Phenomenology, Bombay,
edited by D.P. Roy and P. Roy (World Scientific, Singapore, 1989), p.346;
A. Mishra, H. Mishra, S.P. Misra and S.N. Nayak, Phys. Rev. D44, 110 (1991)}

\def \nmtr {A. Mishra,
H. Mishra and S.P. Misra, Int. J. Mod. Phys. A5, 3391 (1990); H. Mishra,
 S.P. Misra, P.K. Panda and B.K. Parida, Int. J. Mod. Phys. E 1, 405, (1992);
 {\it ibid}, E 2, 547 (1993)}

\def \dtrn {P.K. Panda, R. Sahu and S.P. Misra,
Phys. Rev C45, 2079 (1992)}

\def \qcd {G. K. Savvidy, Phys. Lett. 71B, 133 (1977);
S. G. Matinyan and G. K. Savvidy, Nucl. Phys. B134, 539 (1978); N. K. Nielsen
and P. Olesen, Nucl.  Phys. B144, 376 (1978); T. H. Hansson, K. Johnson,
C. Peterson Phys. Rev. D26, 2069 (1982)}

\def \cornwal {J.M. Cornwall, Phys. Rev. D26, 1453 (1982)}

\def \mndglv {J. E. Mandula and M. Ogilvie, Phys. Lett. 185B, 127 (1987)}

\def \schwinger {J. Schwinger, Phys. Rev. 125, 1043 (1962); ibid,
127, 324 (1962)}

\def \schutte {D. Schutte, Phys. Rev. D31, 810 (1985)}

\def \amspm {A. Mishra and S.P. Misra, Z. Phys. C 58, 325 (1993)}

\def \gft{ For gauge fields in general, see e.g. E.S. Abers and
B.W. Lee, Phys. Rep. 9C, 1 (1973)}

\def \gribov {V.N. Gribov, Nucl. Phys. B139, 1 (1978)}

\def \spmftm {S.P. Misra, Phys. Rev. D18, 1661 (1978); {\it ibid}
D18, 1673 (1978)}

\def \lopr {A. Le Youanc, L.  Oliver, S. Ono, O. Pene and J.C. Raynal,
Phys. Rev. Lett. 54, 506 (1985)}

\def \spphi {S.P. Misra and S. Panda, Pramana (J. Phys.) 27, 523 (1986);
S.P. Misra, {\it Proceedings of the Second Asia-Pacific Physics Conference},
edited by S. Chandrasekhar (World Scientfic, 1987) p. 369}

\def\spmdif {S.P. Misra and L. Maharana, Phys. Rev. D18, 4103 (1978);
    S.P. Misra, A.R. Panda and B.K. Parida, Phys. Rev. Lett. 45, 322 (1980);
    S.P. Misra, A.R. Panda and B.K. Parida, Phys. Rev. D22, 1574 (1980)}

\def \spmvdm {S.P. Misra and L. Maharana, Phys. Rev. D18, 4018 (1978);
     S.P. Misra, L. Maharana and A.R. Panda, Phys. Rev. D22, 2744 (1980);
     L. Maharana,  S.P. Misra and A.R. Panda, Phys. Rev. D26, 1175 (1982)}

\def\spmthr {K. Biswal and S.P. Misra, Phys. Rev. D26, 3020 (1982);
               S.P. Misra, Phys. Rev. D28, 1169 (1983)}

\def \spmstr { S.P. Misra, Phys. Rev. D21, 1231 (1980)}

\def \spmjet {S.P. Misra, A.R. Panda and B.K. Parida, Phys. Rev Lett.
45, 322 (1980); S.P. Misra and A.R. Panda, Phys. Rev. D21, 3094 (1980);
  S.P. Misra, A.R. Panda and B.K. Parida, Phys. Rev. D23, 742 (1981);
  {\it ibid} D25, 2925 (1982)}

\def \arpftm {L. Maharana, A. Nath and A.R. Panda, Mod. Phys. Lett. 7,
2275 (1992)}

\def \van {R. Van Royen and V.F. Weisskopf, Nuov. Cim. 51A, 617 (1965)}

\def \rchpi {S.R. Amendolia {\it et al}, Nucl. Phys. B277, 168 (1986)}

\def \chrl{ Y. Nambu, Phys. Rev. Lett. \zbf 4, 380 (1960);
 J.R. Finger and J.E. Mandula, Nucl. Phys. \zbf B199, 168 (1982);
A. Amer, A. Le Yaouanc, L. Oliver, O. Pene and
J.C. Raynal, Phys. Rev. Lett.\zbf  50, 87 (1983);
ibid, Phys. Rev.\zbf  D28, 1530 (1983); S.L. Adler and A.C. Davis,
Nucl. Phys.\zbf  B244, 469 (1984); R. Alkofer and P. A. Amundsen,
Nucl. Phys.\zbf B306, 305 (1988); M.G. Mitchard, A.C. Davis and A.J.
Macfarlane, Nucl. Phys. \zbf B325, 470 (1989);
B. Haeri and M.B. Haeri, Phys. Rev.\zbf  D43,
3732 (1991); S. Li, R.S. Bhalerao and R.K. Bhaduri, Int. J. Mod. Phys.
\zbf A6, 501 (1991); V. Bernard, Phys. Rev.\zbf  D34, 1601 (1986);
S.P. Klevensky, Rev. Mod. Phys.\zbf  64, 649 (1992); S. Schram and
W. Greiner, Int. J. Mod. Phys. \zbf E1, 73 (1992)}

\def \spmijp { S.P. Misra, Ind. J. Phys. 61B, 287 (1987)}

\def \feynman {R.P. Feynman and A.R. Hibbs, {\it Quantum mechanics and
path integrals}, McGraw Hill, New York (1965)}

\def \glstn{ J. Goldstone, Nuov. Cim. \zbf 19, 154 (1961);
J. Goldstone, A. Salam and S. Weinberg, Phys. Rev. \zbf  127,
965 (1962)}

\def \anderson {P.W. Anderson, Phys. Rev. \zbf {110}, 827 (1958)}

\def \nambu{ Y. Nambu, Phys. Rev. Lett. \zbf 4, 380 (1960)}

\def\donogh {J.F. Donoghue, E. Golowich and B.R. Holstein, {\it Dynamics
of the Standard Model}, Cambridge University Press (1992)}

\def\satz {T. Matsui and H. Satz, Phys. Lett. B178, 416 (1986)}

\def\cps {C. P. Singh, Phys. Rep. 236, 149 (1993)}

\vfil
\end{document}